\documentclass[fleqn,10pt,a4paper]{article}

\usepackage{graphicx,color}	%graphics
\usepackage{soul}
\usepackage{xr}
\usepackage{a4wide}
\usepackage{amsmath}
\usepackage[utf8]{inputenc}

\newcommand{\e}[2]{$#1 \times 10^{#2}$}
\newcommand{\pval}{$P$-value}

\title{Functional transcription factor target discovery via compendia of binding and expression profiles}

\author{Christopher J.\ Banks$^1$, Anagha Joshi$^{2,*}$ and Tom Michoel$^{1,*}$}

\date{}

\begin{document}

\maketitle

\begin{flushleft}
  $^1$ Division of Genetics and Genomics, The Roslin Institute, The University of Edinburgh, UK\\
  $^2$ Division of Developmental Biology, The Roslin Institute, The University of Edinburgh, UK\\[2mm]
  $^*$ Correspondence to: Anagha.Joshi@roslin.ed.ac.uk or Tom.Michoel@roslin.ed.ac.uk
\end{flushleft}

\bigskip

\begin{abstract}
  Genome-wide experiments to map the DNA-binding locations of transcription-associated factors (TFs) have shown that the number of genes bound by a TF far exceeds the number of possible direct target genes. Distinguishing functional from non-functional binding is therefore a major challenge in the study of transcriptional regulation.  We hypothesized that functional targets can be discovered by correlating binding and expression profiles across multiple experimental conditions. To test this hypothesis, we obtained ChIP-seq and RNA-seq data from matching cell types from the human ENCODE resource, considered promoter-proximal and distal cumulative regulatory models to map binding sites to genes, and used a combination of linear and non-linear measures to correlate binding and expression data. We found that a high degree of correlation between a gene's TF-binding and expression profiles was significantly more predictive of the gene being differentially expressed upon knockdown of that TF, compared to using binding sites in the cell type of interest only. Remarkably, TF targets predicted from correlation across a compendium of cell types were also predictive of functional targets in other cell types. Finally, correlation across a time course of ChIP-seq and RNA-seq experiments was also predictive of functional TF targets in that tissue.
\end{abstract}

%%%%%%%%%%%%%%%%%%%%%%%%%%%%%%%%%%%

\section*{Background}
\label{sec:background-tom}

Transcriptional regulation by DNA-binding transcription-associated factors and chromatin modifiers (here collectively abbreviated as ``TFs'') is a fundamental process determining all aspects of cell behaviour, and TFs are known to be essential for a wide range of important cellular and organismal phenotypes. Using ChIP-sequencing technology\cite{park2009}, the ENCODE and modENCODE consortia have generated detailed maps of the genomic locations where TFs bind in diverse human cell types \cite{encode2012} and in important model organisms \cite{gerstein2010integrative, roy2010identification,
  yue2014comparative}. Invariably, these experiments have demonstrated that TFs bind throughout the genome over a quantitative continuum of occupancy levels \cite{biggin2011animal} and that the number of binding events can significantly exceed the number of known or possible direct target genes \cite{macquarrie2011genome}. Hence, predicting when the binding of a TF in a gene locus will lead to a biologically significant change in the rate of transcription of the neighbouring gene (``functional DNA binding'', see the glossary of terms by Biggin \cite{biggin2011animal}) is a challenging and largely unsolved problem. 

Several studies have recently used ChIP-sequencing data of TFs and/or histone modifications to predict absolute expression levels in a particular cell type \cite{ouyang2009chip, Cheng2012bib2, dong2012modeling, budden2014predicting}. While these studies show that a large proportion of the variation in expression levels across genes can be explained by the presence or absence of TF-binding sites for particular combinations of TFs \cite{ouyang2009chip, Cheng2012bib2}, this approach is ill-suited to predict functional TF targets, i.e. to predict \emph{differential} gene expression in a particular cell type upon perturbation of the TF. Indeed, a recent large-scale study where 59 TFs where knocked down in a human lymphoblastoid cell line (GM12878) concluded that only a small subset of genes bound by a factor within 10kb of their transcription start site (TSS) were differentially expressed following knockdown of that factor \cite{cusanovich2014functional}. However, Cheng et al.\ \cite{Cheng2012bib2} also showed that differential TF binding between two cell types correlates with differential gene expression between those two cell types, suggesting that functional TF target genes can possibly be predicted through the ``guilt-by-association'' principle by correlating TF-occupancy and gene expression levels across multiple cell types.

In other applications of genomics, function is often predicted in this manner. Genes with similar expression profiles \cite{eisen1998cluster,hugh00}, genetic interaction profiles \cite{costanzo2011charting} or protein interaction partners \cite{sharan2007network} often share the same molecular function. Likewise, putative DNA-regulatory motifs are identified from their shared occurrence in the upstream regulatory sequences of co-expressed genes \cite{roth1998finding}, networks of TF-regulatory interactions are inferred by associating TF activity profiles to candidate target expression levels \cite{friedman2004,bussemaker2007,balwierz2014ismara}, and long-range DNA contact interactions between regulatory elements and putative target genes can be predicted by correlating open chromatin (measured by sequencing DNase I hypersensitive sites \cite{boyle2008high}) and gene expression levels across multiple cell types \cite{xi2007identification,natarajan2012predicting,
  sheffield2013patterns, marstrand2014identifying, demeyer2014graph}.

To test the hypothesis that the guilt-by-association principle applies to the discovery of functional TF target genes, we used ChIP- and RNA-sequencing data across multiple cell types from the human ENCODE resource \cite{encode2012}.  We considered several cumulative regulatory models to map ChIP-peaks to genes, ranging from proximal binding to incorporating distal events: 1kb/5kb/10kb/50kb around the TSS, the nearest TSS, and 1kb/5kb around the TSS and in the gene body. This is consistent with the emerging view that TF-binding sites act redundantly to promote robustness against genetic and environmental perturbations, and that they may regulate their target genes in a cumulative manner \cite{spivakov2014spurious}.  Furthermore, since there exists no gold standard data of functional DNA-binding events (in the sense defined above), we used the knockdown data of Cusanovich et al.\ \cite{cusanovich2014functional} as a proxy measurement. Five of the knocked down factors had ChIP-seq binding maps available in at least ten cell types in the human ENCODE data, with matching RNA-seq gene expression data, and were considered in this study. Using three different correlation measures, individually and in combination, we found that the correlation between variation in cumulative binding around the TSS of a gene and variation in expression levels of that gene was a better predictor of functional effects than the presence of multiple binding events in the cell type where the TF knockdown was performed. Remarkably, these results were confirmed when using correlation across a time course of ChIP-seq and RNA-seq experiments during mouse circadian rhythm \cite{koike2012transcriptional}.

\section*{Methods}

\subsection*{Preparation of data}

The binding events (peaks) for eight transcription-associated factors (CEBPB, EP300, EZH2, MYC, RAD21, REST, TAF1 and YY1) with binding profiles in ten or more cell types were downloaded from the ENCODE resource \cite{encode2012}. Peaks were mapped to transcription start sites (GENCODE v12) using seven different models (1kb/5kb/10kb/50kb around the TSS, the nearest TSS, and 1kb/5kb around the TSS and in the gene body). To calculate the peak height at each peak, we downloaded the corresponding mapped read files (BED files) for each sample from the ENCODE resource. We then calculated the coverage using BEDTOOLs and normalized coverage count was used as an estimate for peak height. The RPKM values of RNA sequencing data (GENCODE v12) for the corresponding cell lines were also downloaded from the ENCODE resource. They were then quantile normalised using R to be comparable across samples. ENCODE binding and expression profiles were available for 24,392 genes. The global expression change (in the form of differential expression \pval s) upon deletion of five of the eight factors (EP300, EZH2, RAD21, TAF1 and YY1) in a lymphoblastoid cell line (GM12878) was available for 8,872 genes (also called the ``reference genes'' below) \cite{cusanovich2014functional}. Differentially expressed genes upon deletion of CEBPB were obtained directly from the Gene Expression Omnibus using the GEO2R tool (accession number: GSE54975). Differentially expressed genes upon deletion of MYC were obtained from Seitz et al.\ \cite{seitz2011}.

The binding events (peaks) for six circadian regulators (Bmal1, Clock, Cry1, Cry2, Per1, Per2) and two RNA polymerase II states (8WG16 and ser5p) at six time points (1hr, 4hr, 8hr, 12hr, 16hr and 20hr) as well as the RPKM values of RNA sequencing data at the same time points from murine liver samples were obtained from Koike et al.\ \cite{koike2012transcriptional}. A total of 2629 genes had both binding and RPKM data available for Per2. Differentially expressed genes upon deletion of Per2 in murine liver \cite{zani2013per2} were obtained directly from the Gene Expression Omnibus using the GEO2R tool (accession number GSE30139; 2409 genes with fold change $>$1.5 and FDR corrected \pval$<$0.05).

From each source of data, we obtained a number of working sets of data where for each TF there exists concordant data from binding, expression, and knockdown.  For each TF and each peak-to-gene model there were three datasets. The ChIP and expression data were matrices with a row for each gene and a column for each condition, containing respectively the number of peaks mapped to the TSS of that gene and its expression level. The knockdown data for ENCODE was an expression change \pval\ for each gene and for mouse circadian was a list of genes with significant expression change.

We also prepared ChIP data for the same factors with quantitative binding information (sum of peak magnitudes for each gene). Finally we prepared alternative datasets filtered for CpG-rich and CpG-depleted promoters, as obtained from the UCSC genome browser.

\subsection*{Prediction of functional TF target genes}
The approach we took to predicting functional target genes looked at the correlation between binding and expression profiles over a range of cell-types or conditions. However, we found that correlation by any one method alone is a not necessarily a good predictor for any given factor or binding model. We found that different correlation methods identify different types of relation between binding and expression. To improve prediction we combined results from a number of correlation methods in a wisdom of crowds approach. 

For each dataset (i.e. for each TF and each peak-to-gene model) we computed the correlation between the number of binding peaks and the expression of each gene across all conditions, using three correlation measures: the absolute Pearson correlation coefficient (PC), the absolute Spearman correlation coefficient (SC), and the absolute combined angle ratio statistic (CARS). CARS is our variant of the angle ratio statistic (ARS) of Marstrand and Storey \cite{marstrand2014identifying}. The ARS was shown to have high power for detecting associations when both variables are restricted to a narrow relative range, with one or very few cell types appearing as distinct outliers \cite{marstrand2014identifying}. Whilst ARS only considers positive associations between variables (i.e. where an increased number of binding sites corresponds to increased expression of a target gene), CARS uses the same principle to test for both positive and negative associations. 
We define CARS as follows. Vector $\vec{x}$ is a vector of RNA-seq data and $\vec{y}$ is a vector of ChIP-seq data (for the same cell-types). Both vectors have length $t$. Both vectors are then scaled: $\vec{x^s} = \frac{\vec{x}}{\max(|\vec{x}|)}$ and likewise for $\vec{y^s}$. Both vectors are then median centred: $\vec{x^*} = \vec{x^s} - \mathrm{med}(\vec{x^s})$ and likewise for $\vec{y^*}$. Outlier distance is measured for each point:
\begin{equation}
  \vec{d} = \left(\sqrt{{x^*_1}^2 + {y^*_1}^2},\ldots,\sqrt{{x^*_t}^2 + {y^*_t}^2}\right)
\end{equation}
and the ratio statistic
\begin{equation}
  \vec{r} = \left(\frac{d_1}{\mathrm{med}(\vec{d})},\ldots,\frac{d_t}{\mathrm{med}(\vec{d})}\right) 
\end{equation}
quantifies the distance of each point from the medoid.  We then take the angle of each point from the x-axis $\vec{\theta} = (\theta_1,\ldots,\theta_t)$ and form a positive angle statistic $\vec{\Delta^+} = (\Delta^+_1,\ldots,\Delta^+_t)$, the angular distance of each point from the line $x=y$, and a negative angle statistic $\vec{\Delta^-} = (\Delta^-_1,\ldots,\Delta^-_t)$, the angular distance of each point from the line $x=-y$: 
\begin{align}
  \Delta^+_i &= \left\{ 
  \begin{array}{l l}
    |45-\theta_i| & 0\leq\theta_i<135 \\
    |225 - \theta_i| & 135\leq\theta_i<315 \\
    |45-(\theta_i-360)| & 315\leq\theta_i<360
  \end{array} \right.,\\
% \hspace{20pt}
  \Delta^-_i = 90- \Delta^+_i &= \left\{ 
  \begin{array}{l l}
    |315-(\theta_i+360)| & 0\leq\theta_i<45 \\
    |135-\theta_i| & 45\leq\theta_i<225 \\
    |315-\theta_i| & 225\leq\theta_i<360
  \end{array} \right..
\end{align}
This is where CARS differs from ARS, which only measures angular distance from the line $x=y$. The positive scores for each point are $\mathit{ARS}^+ = (\mathit{ARS}^+_1,\ldots,\mathit{ARS}^+_t)$ where $\mathit{ARS}^+_i = r_i \times e^{c\Delta^+_i}$ and the negative scores $\mathit{ARS}^-$ are formed in the corresponding manner, with $c<0$ a fixed parameter of the method. The overall combined angle ratio statistic is then defined as
\begin{equation}
  \mathit{CARS} = \left\{ 
  \begin{array}{l l}
    \mathit{ARS}^+_{\max} & \mathit{ARS}^+_{\max} \geq \mathit{ARS}^-_{\max} \\
    \\
    -\mathit{ARS}^-_{\max} & \mathit{ARS}^+_{\max} < \mathit{ARS}^-_{\max}  
  \end{array} \right.
\end{equation}
where $\mathit{ARS}^\pm_{\max}=\max(\mathit{ARS}^\pm)$. The value of the parameter $c$ was determined by requiring that empirical \pval s satisfied a correct null distribution (i.e. such that \pval s\ $>0.5$ had a uniform distribution), following the procedure of Marstrand and Storey \cite{marstrand2014identifying}. In this study we used $c=-0.01$ which conformed to this requirement for all data sets.

For all three correlation measures, we kept track of whether the score originated from a positive or negative association.  For comparison purposes, we also considered the number of binding peaks in the cell type where the knockdown experiment was performed (called ``multiple binding'') as a functional target predictor.

\subsection*{Measures of performance}
We used the knockdown data as our gold standard for defining a functional effect of TF-binding on target gene expression.  We measured the predictive performance of each correlation method and multiple binding by computing the precision vs.\ recall (PR) curves (where \emph{precision} is the proportion of predicted genes that are in the gold standard and \emph{recall} is the proportion of the gold standard set that was predicted) for each dataset.  We then took, for each dataset, thresholds on both the level of multiple binding and the correlation scores, and genes with scores above the threshold were taken as positive predictions. To achieve comparable results between different TFs, we chose thresholds to give a specific fold increase in precision over the background proportion of bound genes with a knockdown effect (i.e., the proportion of genes bound in the knockdown cell type that have a significant functional effect in the knockdown). We recorded the positively identified genes and calculated the hypergeometric overlap \pval\ between the predicted gene set and gold standard set. We chose to make comparisons at a 1.5-fold precision increase over the background. In other words, if a fraction $f$ of the reference gene set were differentially expressed upon knockdown of a particular TF, we determined thresholds for the PC, SC and CARS such that a fraction $1.5f$ of the genes exceeding the threshold were differentially expressed.  For making predictions for TFs for which no gold standard set was available, we chose the threshold to be the top 1\% of predictions.

Performance was measured on the intersection of genes with available knockdown differential expression data (the reference gene set) and real correlation scores (i.e., non-constant binding and expression profiles) or multiple-binding score (i.e., at least one peak in the knockdown cell type for the current peak-to-gene model); in the CARS method, outliers in one dimension (e.g. with constant binding profile) were penalized by the angular penalty but not excluded from the calculation, as recommended by  Marstrand and Storey \cite{marstrand2014identifying} to ensure a correct null distribution. For the biological validation of the predicted target sets, all genes in the complete set of 24,392 genes above the given score threshold were used, irrespective of the availability of knockdown data.

\section*{Results}
\subsection*{Number of binding sites in a gene locus is a weak predictor of functional relevance of binding}

In order to test the hypothesis that correlation between TF binding and gene expression across cell types can be used as a predictor of the functional relevance of binding events, we used data from the human ENCODE resource \cite{encode2012}. We selected eight TFs, each with ChIP-sequencing profiles in at least ten cell lines and corresponding RNA-sequencing data available for the same cell lines. We first assigned genome-wide binding locations (peaks) to putative target genes if they were within 5kb of the transcription start sites (TSSs) obtained from GENCODE V12.  We then defined the binding profile of a gene as the number of peaks associated to that gene across the available cell types. We defined functional relevance of binding of a TF in a gene locus by whether or not the gene is differentially expressed upon knockdown of the transcription factor. Although a limited and stringent definition of functional relevance, this facilitated us to make a systematic and quantitative comparison of different functional prediction methods. For five of the selected TFs (EP300, EZH2, RAD21, TAF1 and YY1), siRNA knockdown data in a lymphoblastoid cell line (GM12878) was available in the form of differential expression $P$-values for 8,872 genes \cite{cusanovich2014functional}. The differentially expressed genes (\pval$<$0.05) were considered functional targets of the corresponding transcription factor (true positives) and the non-differentially expressed genes true negatives. This resulted in five datasets (one for each TF) with binding profiles for 24,392 genes, together with matching expression profiles for the same genes in the same cell types as well as true positive and true negative functional target gene lists (see Methods for details).

We first calculated for each TF the ratio of functional targets among genes bound by the TF in GM12878.  These ratios showed only a very limited increase compared to the genome-wide background ratio of functional targets, which was significant for only two factors (EP300 and RAD21, hypergeometric \pval\ $<0.05$, Supplementary Table 1). %\ref{table:summary1}
This result is consistent with the finding by Cusanovich et al.\ \cite{cusanovich2014functional} that only 12 of their 29 knockdowns with available ChIP-sequencing data resulted in a significant overlap between binding and differential expression.  We then tried to improve results by predicting targets only if multiple binding sites are present. For three factors (EP300, RAD21, and YY1), we could achieve a 1.5-fold increase in precision (i.e., percentage of known functional targets) with hypergeometric \pval\ $<0.05$. However, the threshold number of binding sites had to be large and consequently the predicted target sets were small (Supplementary Table 1). %\ref{table:summary1}).

\subsection*{Correlation across compendia of binding and expression profiles improves functional target prediction}

\begin{figure}[h!]
  \centering
  \includegraphics[width=.99\linewidth]{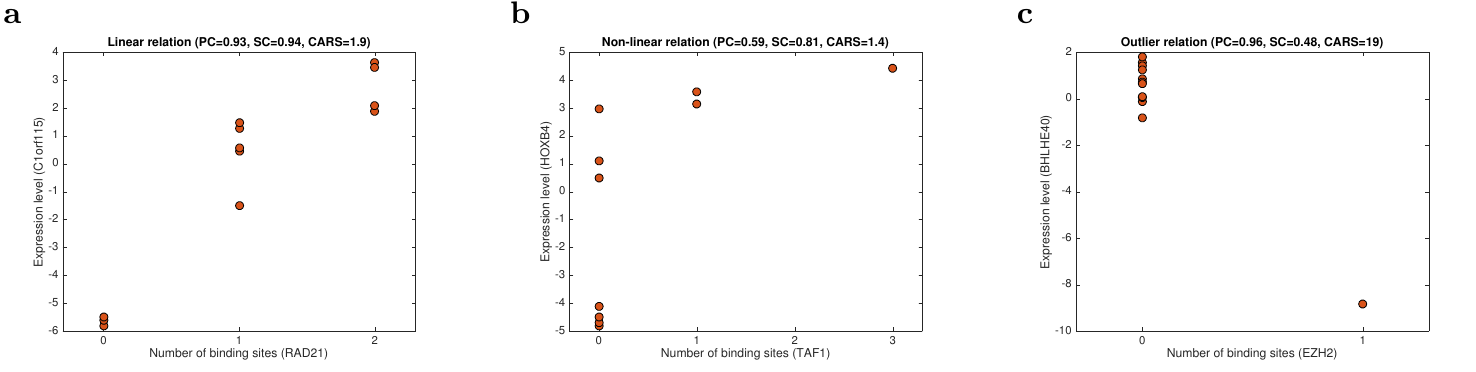}
  \caption{Characteristic scatter plots of binding and expression profiles for known differentially expressed targets, showing a linear relation favoured by Pearson correlation (\textbf{a}), a non-linear monotonic relation favoured by Spearman correlation (\textbf{b}), and an outlier relation favoured by CARS (\textbf{c}).}
  \label{fig:profiles}
\end{figure}

To investigate the utility of correlation-based methods to predict functional targets, we considered three distinct association measures between binding and expression profiles across multiple cell types: the absolute Pearson correlation coefficient (PC), which tests for a positive or negative linear association (Figure \ref{fig:profiles}a), the absolute Spearman correlation coefficient (SC), which tests for a positive or negative monotonic trend (Figure \ref{fig:profiles}b), and the combined angle ratio test statistic (CARS), which tests for a non-linear ``on-off'' relationship in a positive or negative direction (Figure \ref{fig:profiles}c) (see Methods).

\begin{figure}
  \centerline{\includegraphics[width=1.2\linewidth]{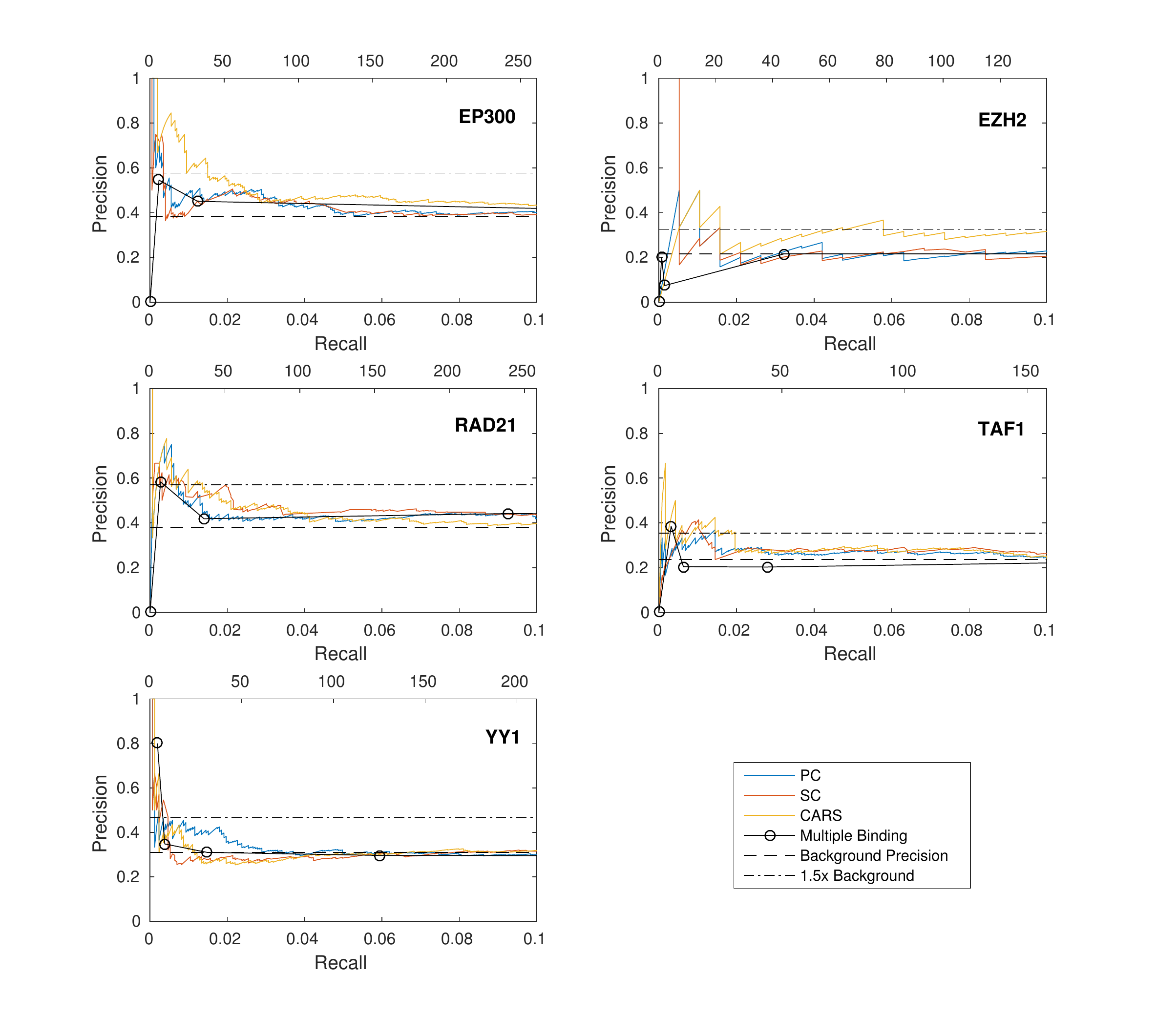}}
  \caption{Precision vs. Recall curves for functional binding of EP300, EZH2, RAD21, TAF1 and YY1 in the 5kbTSS peak to gene model predicted by each method, using 8,872 reference genes (true positive and true negative functional targets). The top scale on the x-axis shows the number of true positives for the corresponding recall value.}
  \label{fig:PRs}
\end{figure}

\begin{figure}
  \centerline{\includegraphics[width=\linewidth]{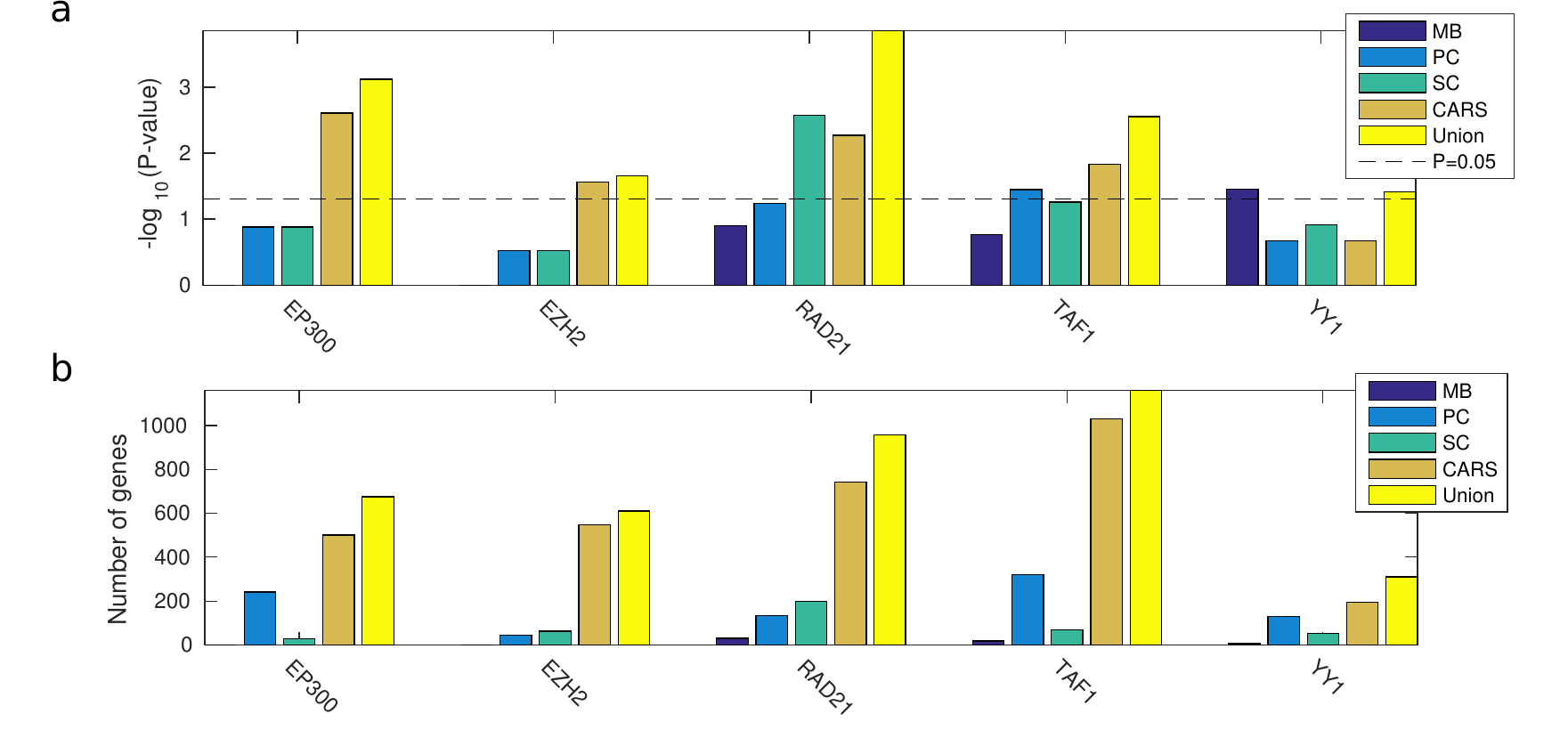}}
  \caption{Functional target set enrichment significance (negative log hypergeometric \pval) (\textbf{a}) and size (\textbf{b}) for each method and TF at a predicted 1.5-fold precision over background. This is compared to using binding data from the knockdown cell type only (MB). Significance was determined by analysing the 8,872 gold standard reference genes, whereas sets sizes refer to subsets of the complete set of 24,392 genes with correlation score exceeding the 1.5-fold precision threshold derived from the gold standard.}
  \label{fig:sigLevels}
\end{figure}

We validated the predictive performance of each method for each TF independently using precision-recall (PR) curves. In all cases, high-scoring genes for any of the correlation measures were more likely to be differentially expressed upon knockdown of the TF than low-scoring genes (Figure~\ref{fig:PRs}). In particular, for four out of five factors (EP300, EZH2, RAD21, YY1) a small number of targets were predicted with a precision close to one.  This enrichment (high-precision-low-recall region) translated into an improvement in the number of significant sets and the size of set predicted compared to using multiple binding in the knockdown cell type. To enable comparison of different methods across different factors, we determined for each method and each factor the score threshold resulting in a 1.5-fold increase in precision compared to the genome-wide background ratio of functional targets, and calculated the significance of the overlap between all functional targets and targets predicted at this threshold using a hypergeometric test. We found significantly enriched target sets for each factor for at least one, and often multiple, of the methods, predicting significantly more target genes compared to using binding data from the knockdown cell type only (MB) (Figure~\ref{fig:sigLevels} and Supplementary Table 2).%\ref{table:summary2}).

\begin{figure}
  \centerline{\includegraphics[width=\linewidth]{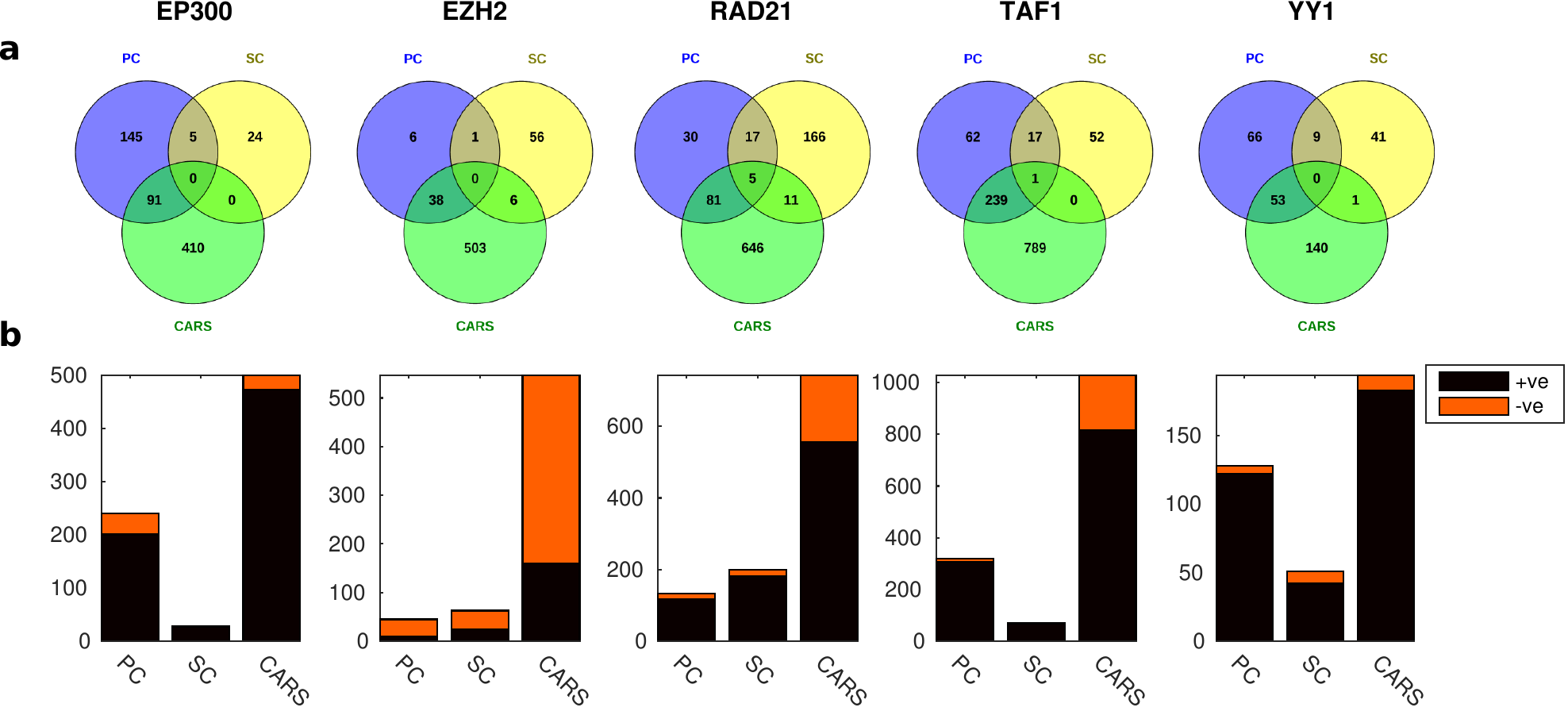}}
    \caption{Overlap of functional target sets for each factor predicted by each method at a predicted 1.5-fold precision over background (\textbf{a}). Number of positively and negatively correlated targets predicted by each method for each factor at a predicted 1.5-fold precision over background (\textbf{b}). Set sizes in both panels refer to subsets of the complete set of 24,392 genes exceeding the 1.5-fold precision threshold determined by analysing the 8,872 reference genes.}
  \label{fig:venns}
\end{figure}

Interestingly, taking the union of the predicted gene sets for each method gave a further increase in overlap with the gold standard (Figure~\ref{fig:sigLevels} and Supplementary Table 2). %\ref{table:summary2}). 
This was explained by the fact that enriched target sets predicted by each method (PC, SC, and CARS) showed only a limited overlap (Figure \ref{fig:venns}a), suggesting that all types of relations (linear, non-linear monotonic, on-off; cf. Figure \ref{fig:profiles}) do occur between binding and expression profiles of functional TF targets. 

Since all three methods take into account both positive and negative associations between binding and expression, we asked whether specific TFs show a bias for either sign. For four out of five factors (EP300, RAD21, TAF1 and YY1), positive interactions dominated the predicted target sets, suggesting that they mostly function as activators of expression. In contrast, for EZH2 most predicted interactions were negative (Figure \ref{fig:venns}b), consistent with the fact that more than two thirds of the differentially expressed genes were up-regulated in response to EZH2 knockdown \cite{cusanovich2014functional}. EZH2 is indeed known to repress transcription by participating in histone mark H3K27me3 deposition as well as DNA methylation \cite{vire2006polycomb}, although it also functions as a double-faceted molecule in breast cancers, either as a transcriptional activator or repressor of NF-kB targets, depending on the cellular context \cite{Lee2011798}.

Cell-type specific studies have previously used quantitative TF binding information (i.e. peak heights) to predict absolute expression levels \cite{ouyang2009chip, Cheng2012bib2}. We therefore also tested our method using ChIP-seq peak heights in place of counts. However we found no significant improvement in prediction using quantitative information. Precision levels varied, but were similar to results obtained via on-off binding (see Supplementary Figures 3,4 %\ref{fig:sig5kbPH},\ref{fig:sigAllPH}
---compare with Figure 3). %\ref{fig:sigLevels}). 
Cheng et al.\ \cite{Cheng2012bib2} also found that the cell-type specific expression of CpG-rich promoters was easier to predict than CpG-depleted promoters. We therefore partitioned the gene set in the same way, and compared results. %\ref{fig:CpGresults}). 
Here too, the results showed that there is no significant difference on the precision overall between CpG-rich and CpG-depleted genes  (see Supplementary Figure 5), even after correcting for the difference in the number of CpG-rich and CpG-depleted genes (see Supplementary Figure 6). However some improvement is shown for EZH2; this is not unreasonable, given that EZH2 is known to target CpG-island promoters \cite{riising2014gene}.

To understand the functional relevance of the predicted target sets, we performed gene ontology enrichment analysis using BiNGO \cite{maere2005bingo}. The target gene sets of each factor (Supplementary Data 1) were enriched for distinct and specific functional categories, such as ``inflammatory response'' (EP300; \pval\ $<$ \e{1.2}{-7}), ``generation of neurons'' (EZH2; \pval\ $<$ \e{1.6}{-12}). In contrast, the sets of genes bound a factor in the knockdown cell type (Supplementary Data 2) were enriched for metabolic processes for all factors. Furthermore, of the 2,618 predicted functional target genes across all five TFs, 70\% were predicted to be target of only one factor. In contrast, 82\% of the 5,764 bound genes were bound by more than one TF. Taken together, these results suggest that the correlation-based method is able to select smaller and more specific sets of functional target genes from the hundreds to thousands of genes bound by a given factor in a given cell type.

Full target sets are available in Supplementary Datasets 59--93.

\subsection*{Correlation between binding and expression predicts functional targets in non-ENCODE cell types}
\label{sec:bett-pred-meth}
In the previous analyses, the gold standard validation data (differential expression results) were obtained in a cell type that was also present as one of the ENCODE cell lines used to predict targets. Next we asked whether correlation of binding and expression across a compendium of cell types is also informative for predicting targets in cell types not present in the compendium. 

Firstly, we predicted functional targets for three additional transcription factors (MYC, CEBPB and REST) where binding and expression information was available in ten or more ENCODE cell types, but not the perturbation data. These TFs typically bind to a few thousand gene loci in any given cell type, but the correlation-based method enabled us to select only a few hundred high-confidence functional targets by taking the union of top 1\% predictions from each method.  The predicted targets of CCAAT/enhancer binding protein beta (CEBPB) were specifically enriched for the Wnt signalling pathway (\pval\ = \e{7.9}{-4}). CEBPB has a demonstrated role in the suppression of Wnt/$\beta$-catenin signaling during adipogenesis \cite{chung2012regulation}.  CEBPB targets were also enriched for the functional category ``positive regulation of cytokine production during immune response'' (\pval\ = \e{9.1}{-2}), in line with the well characterised role of CEBPB in the regulation of immune and inflammatory response genes \cite{pruitt2014refseq}. Since the ENCODE cell types are cancer-related rather than immunological or adipogenesis-related, this demonstrates the strength of the correlation-based approach to find the most functionally relevant targets of a transcription factor. Similarly, the predicted targets of REST were enriched for functions specific to neurons, such as presynaptic membrane (\pval\ = \e{4.1}{-3}). REST is identified as a key regulator to protect neurons in parts of the brain from oxidative stress, as well as protein aggregations characteristic of many neurodegenerative diseases \cite{lu2014rest}.

Next, having confirmed that the correlation-based method is able to identify the most functionally relevant targets, we investigated whether the high-confidence predicted targets are also valid in other cell types. Reschen et al.\ \cite{reschen2015lipid} systematically investigated the role of CEBPB in macrophage differentiation in an \textit{in vitro} model for coronary artery disease. By performing ChIP sequencing for CEBPB before and after macrophages differentiate into foam cells, they identified 5866 genes where CEBPB was bound at significantly higher levels in foam cells. Of these differentially bound genes, 16\% (935) were differentially expressed between foam cells and macrophages. Of the 749 predicted CEBPB targets using our correlation-based method across ten ENCODE cell lines, also 16\% (119) of genes were differentially expressed between foam cells and macrophages. Similarly, Seitz et al.\ \cite{seitz2011} studied the role of MYC in Burkitt Lymphoma (BL) and identified 7054 MYC binding sites (6169 within 5kb of a TSS) in 5 BL cell lines. 530 (8.5\%) of these bound genes were differentially expressed after siRNA-mediated knock-downs of MYC in BL cell lines.  Our method predicted 728 MYC targets using the ENCODE data, which showed 9\% overlap with genes differentially expressed after siRNA-mediated knock-downs of MYC in BL cell lines. These two analyses demonstrate that predictions derived using our approach have a precision that is comparable to the ChIP sequencing performed in the exact cell line of interest. The recall on the other hand is limited at this point due to the limited availability of cell types with matching binding and expression data to build functional target predictions, and the possibility of missing cell type specific targets.

Full target sets are available in Supplementary Datasets 3--58.

\subsection*{Different binding target models work better for different TFs}
\begin{figure}
  \centerline{\includegraphics[width=\linewidth]{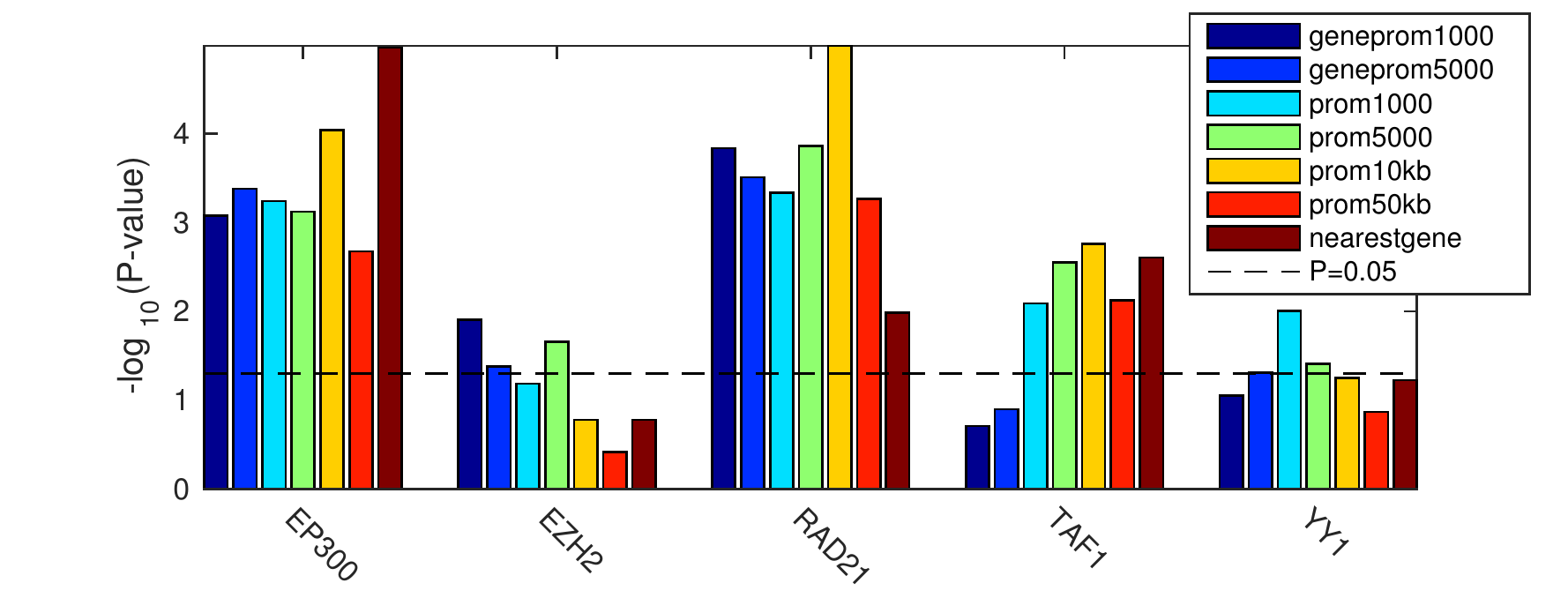}}
  \caption{Functional target set significance (hypergeometric P-value) predicted by the union of all correlation methods for all peak-to-gene models at a predicted 1.5-fold precision over background. Set sizes refer to subsets of the complete set of 24,392 genes exceeding the 1.5-fold precision threshold determined by analysing the 8,872 reference genes.}
  \label{fig:sigPerModel}
\end{figure}

To investigate whether assigning genome-wide binding locations (peaks) to putative target genes using different peak-to-gene models results in improved performance, we considered seven peak-to-gene models: 1kb, 5kb, 10kb or 50kb around the transcription start sites (TSSs) obtained from GENCODE V12; 1kb or 5kb around the TSS and within the gene body; or the nearest TSS. We then predicted the targets using the union of correlation methods for each of the gene models. We noted that different peak-to-gene association models performed best for different factors (Figure \ref{fig:sigPerModel}). Promoter proximal binding has previously been associated with the functional relevance of binding \cite{whitfield2012functional}. Accordingly, for all five TFs, the 5kbTSS correlation gave significant enrichment of functional targets (Figure \ref{fig:sigPerModel}). EZH2 and YY1 performed best when TSS proximal peaks were considered. This suggests that EZH2 and YY1 are involved in transcriptional control through promoter proximal elements only. To note, both EZH2 and YY1 predominantly bind in promoter proximal regions \cite{encode2012}. Given that overall the 5kbTSS model performed better than the 1kbTSS model, this suggests the presence of alternate cell type specific promoters within 5kb of the consensus annotated promoter. For EP300, the farther the peaks from the TSS were considered, the better the predictions, such that the best results were obtained when all peaks were associated to the nearest TSS (Figure \ref{fig:sigPerModel}). As the presence of EP300 is associated with active enhancers \cite{visel2009chip}, it is not surprising to find that the nearest TSS model works best for EP300. Similarly, the 10kb peak-to-gene model was the best performer for RAD21. 

Among the three correlation measures, CARS predicted the highest number of targets at the 1.5-fold precision threshold for most peak-to-gene models, with the exception of the nearest gene model for EP300, where PC performed significantly better, indicating that linear interactions dominate in this case (Supplementary Figures 1 and 2).

\subsection*{Correlation across time points of a defined biological process is also predictive of  functional effects}

\begin{figure}
  \centering
  \includegraphics[width=0.5\linewidth]{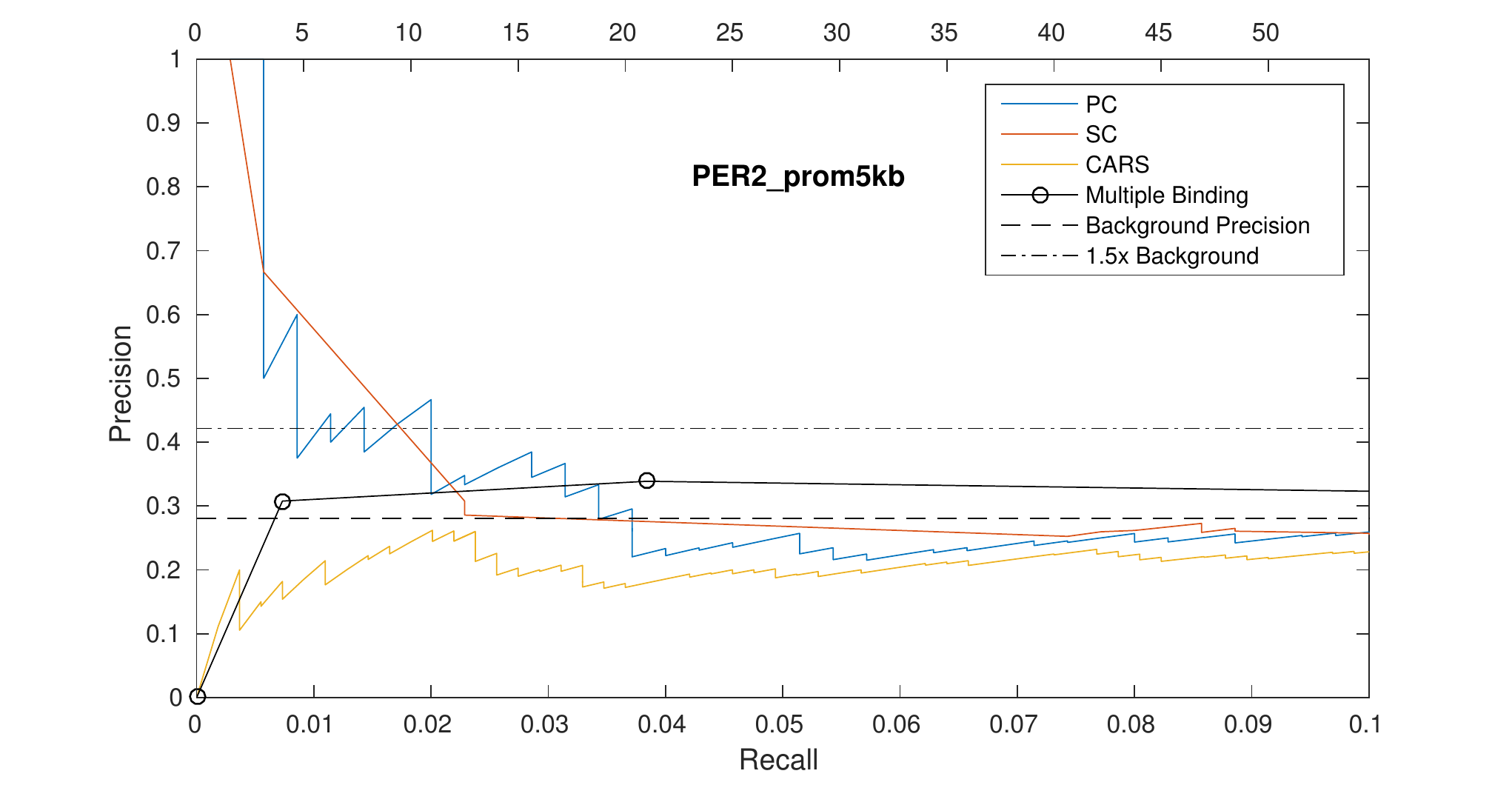}
  \caption{Precision vs. Recall curves for functional binding of PER2 in the 5kbTSS peak-to-gene model by each method, using 21200 reference genes (true positive and true negative functional targets).}
  \label{fig:pr-per2}
\end{figure}

We have demonstrated that correlation between binding patterns and expression values of genes across multiple cell types can be used to predict functional targets. We then investigated whether the correlation-based approach can be extended to other data types, such as time series. The mammalian circadian clock is a cell-autonomous process with a period of about 24 hours. It controls the sleep-wake cycle, blood pressure and hormone secretion by regulating key processes such as metabolism, the cell cycle and DNA repair through feedback loops of transcriptional regulators (Clock, Bmal1, Cry1, Cry2, Per1 and Per2) essential for the rhythmicity \cite{lowrey2011genetics}. We obtained genome-wide binding patterns (ChIP-sequencing data) for the six transcription factors listed above, as well as gene expression profiles (RNA-sequencing data) at six time points (1, 4, 8, 12, 16 and 20 hours) in murine liver \cite{koike2012transcriptional}. We predicted high confidence correlated genes by considering the top 1\% predictions of the union of correlations between binding and expression across this time series, using the 5kbTSS promoter proximal peak-to-gene model. A gold standard set of genes differentially expressed upon TF-perturbation in liver was available only for Per2 \cite{zani2013per2}. As before, the PR curve showed that targets predicted by correlation, but not multiple binding, were enriched for known functional targets (Figure \ref{fig:pr-per2}). 

The high-confidence predicted Per2 targets were highly expressed at 16 and 20 hours, similar to the gene expression pattern of Per2 itself, and were more likely to be bound by Cry1/Cry2 than Clock/Bmal1 \cite{koike2012transcriptional}. They were enriched for the functional category ``regulation of RNA metabolic process'' (\pval\ = \e{4.3}{-3}). Although 10\% of the liver transcriptome follows a circadian rhythm, only about half of it can be explained by de novo transcription, suggesting that mRNA processing may play a key role in the circadian rhythmicity. Per2 is associated with RNA processing through the RNA-dependent methylation process \cite{fustin2013rna}. This again demonstrates that the correlation-based approach enables the identification of a small set of highly-relevant functional targets among the tens of thousands of genes bound by a given transcription factor.

\section*{Discussion}

In this study, we have applied the guilt-by-association principle to predict functional  targets of transcription-associated factors by testing if a gene's TF-specific binding profile across multiple cell types correlated (positively or negatively) with its expression profile across the same cell types, using three distinct correlation measures (Pearson and Spearman correlation and the combined angle ratio statistic) and a range of cumulative regulatory models for mapping TF-binding peaks to transcription start sites. Compared to the traditional approach where target genes are inferred from the presence of one or more binding sites in a gene locus in a cell type of interest, the three correlation-based methods showed improved prediction of functional targets, defined here as genes differentially expressed upon TF knockdown, especially when used in combination.

It is known that TFs function in a condition-specific manner, and it may not be obvious that correlation-based measures across multiple cell types are able to identify functional targets. However, it is precisely the presence of binding and associated change in a target gene's expression level in the cell type(s) where the TF is active, and the absence of this signal in other cell types, which leads to a high-confidence prediction. The angle ratio statistic was developed precisely to detect such cell-type specific effects with high specificity \cite{marstrand2014identifying}, and was indeed found to predict a significantly higher number of functional targets at the same enrichment threshold compared to the Pearson and Spearman correlation, which predominantly select linear or monotonic trends, respectively. Furthermore, if binding of a factor varies across a compendium of cell types, then the correlation between binding and expression was found to be predictive of functional effects even in cell types that were not part of the compendium.  Although more work is needed to investigate the condition-specific properties of predicted functional target genes, our results suggest that correlation-based predictions capture both condition-specific and condition-independent targets of a TF.

Interestingly, these results were confirmed using a time course of matching binding and expression data in a single cell type, showing the wide validity of the guilt-by-association principle for functional TF-binding prediction. In contrast to the ENCODE results, only the Pearson and Spearman correlation predicted significantly enriched target sets in this case, whereas the CARS outlier detection method did not perform well. This is consistent with the fact that samples from the same tissue at different time points are more similar to each other than samples from highly distinct cell types, and emphasizes the importance of combining different correlation methods to detect all types of signal present in a dataset.

Four limitations affect the current study and need to be taken into account. Firstly, our definition of a gold standard of true positive and true negative functional target genes from differential expression data following knockdown of a TF is only a proxy for true functional binding events, namely when the binding of a TF in a gene locus significantly affects the gene's rate of transcription. However no large-scale data of changes in transcription rates following TF knockdowns is currently available. Secondly, although the human ENCODE ChIP-seq matrix currently reports data for nearly 200 TFs and more than 80 cell types, it is very sparse. Indeed, only eight TFs had ChIP-seq profiles available in more than ten cell types with matching RNA-seq data, which we considered a minimum to perform a correlation-based analysis. Of the eight factors considered, half were sequence-specific TFs (CEBPB, MYC, REST and YY1) and half were general factors: two promoter-associated (EZH2 and TAF1), one enhancer-associated (EP300) and one involved in three-dimensional DNA organization (RAD21). Of the sequence-specific TFs, only one (YY1) had knock-out data available in the lymphoblastoid cell line and could thus be validated directly. As more data will become available, it will be important to establish if the reported results also hold for a wider range of sequence-specific transcription factors.  Thirdly, predicting the effect of a particular TF on a particular gene naturally depends on the reliability of the ChIP-seq experiments for that TF, but even within the ENCODE resource, with its high standards for technical quality control, the biological quality of samples is not always guaranteed \cite{devailly2015variable}. Lastly, we only considered the presence of binding sites and expression data to investigate the improvement in functional prediction by correlation-based methods compared to using the presence of binding sites only. Although we found that taking into account binding peak height or promoter CpG content did not improve our predictions, we did not consider the presence of sequence motifs or various chromatin features. These have been shown to improve prediction of cell-type specific variation in expression among genes \cite{ouyang2009chip, Cheng2012bib2}, and will likely also improve prediction of functional targets. Having established the validity of the correlation-based method, future work will be aimed at building functional target gene predictors that combine this approach with additional data types and existing knowledge.  Despite these limitations, we believe that the use of correlated features in compendia of binding and expression profiles with matching conditions is a powerful novel method to predict functional TF target genes, which is able to identify high-confidence functional target genes among the thousands of genes bound by a given transcription factor.

\section*{Acknowledgement}
A.J.\ is a Chancellor’s Fellow of the University of Edinburgh. This work was supported by Roslin Institute Strategic Grant funding from the BBSRC.

\paragraph{Competing interests:} The authors declare no competing interests.
\paragraph{Author contributions:} C.B.\ performed the analysis and wrote the manuscript. A.J.\ and T.M.\ conceived the idea, collected the data, performed the analysis and wrote the manuscript. 

%\bibliography{bibsysbiol,bib2}

\newpage

\appendix

\setcounter{equation}{0}
\renewcommand{\theequation}{S\arabic{equation}}
\setcounter{figure}{0}
\renewcommand{\thefigure}{S\arabic{figure}}
\setcounter{table}{0}
\renewcommand{\thetable}{S\arabic{table}}

\section*{Supplementary figures and tables}
\label{sec:suppl-figur-tabl}

\begin{figure}[h!]
  \centerline{\includegraphics[width=1.4\linewidth]{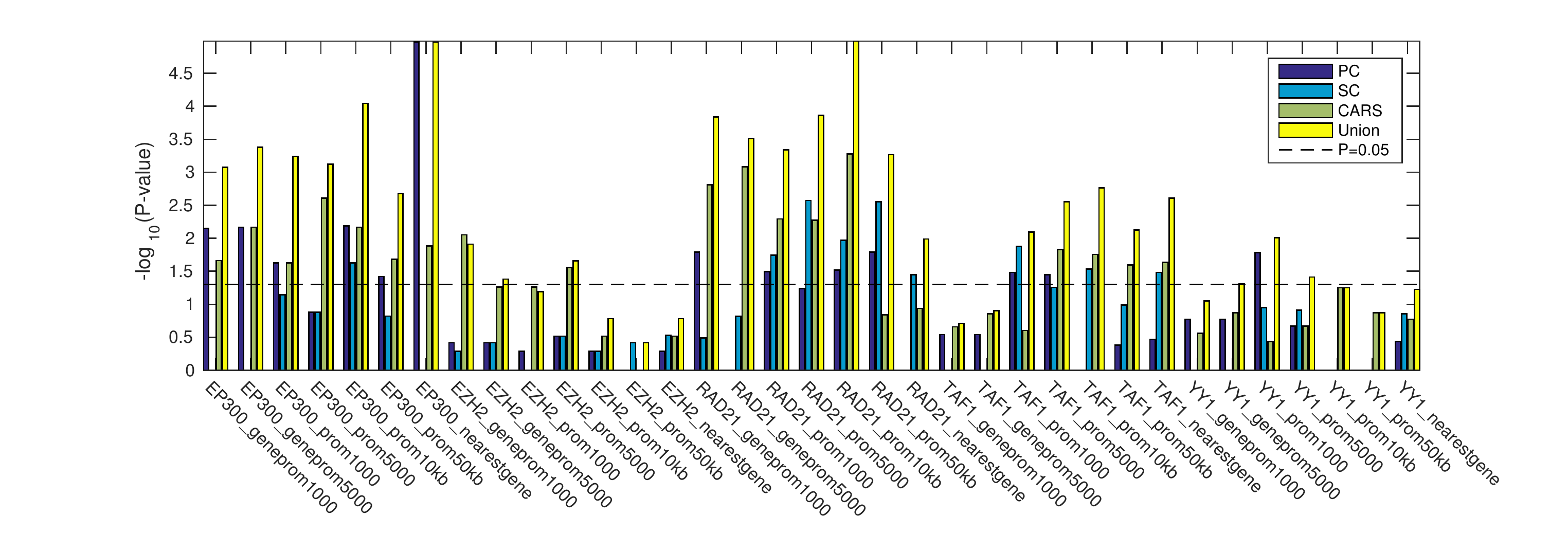}}
  \caption{Functional target set significance (hypergeometric P-value) predicted by each of the correlation methods for all peak-to-gene models at a predicted 1.5-fold precision over background.}
  \label{fig:sigAll}
\end{figure}

\begin{figure}[h!]
  \centerline{\includegraphics[width=1.4\linewidth]{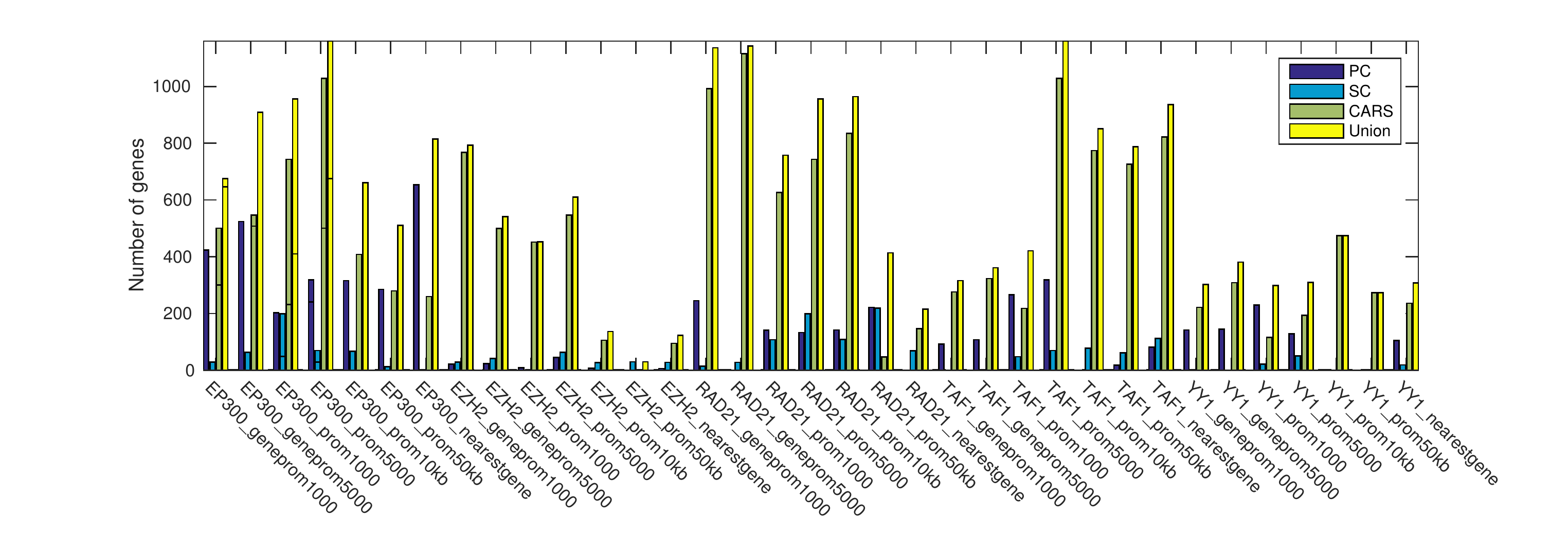}}
  \caption{Functional target set sizes predicted by each of the correlation methods for all peak-to-gene models at a predicted 1.5-fold precision over background. Set sizes refer to subsets of the complete set of 24,392 genes exceeding the 1.5-fold precision threshold determined by analysing the 8,872 reference genes.}
  \label{fig:sizeAll}
\end{figure}

%% Peak heights:
\begin{figure}[h!]
  \centerline{\includegraphics[width=1.2\linewidth]{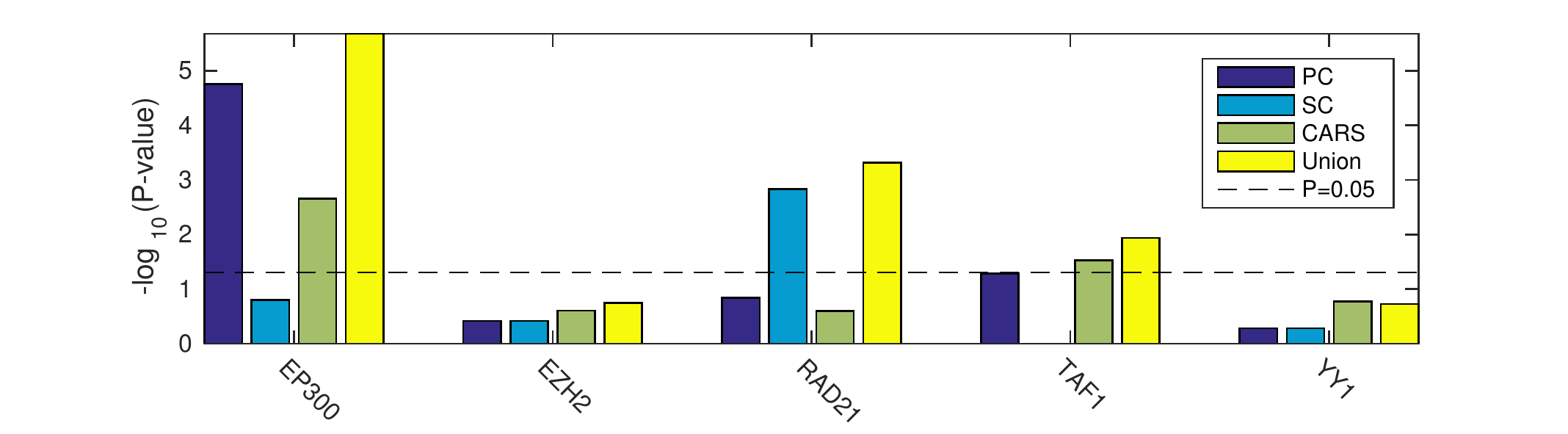}}
  \caption{Functional target set significance (hypergeometric P-value) using quantitative (sum of peak heights) ChIP-seq data predicted by each of the correlation methods for the prom5kb peak-to-gene model at a predicted 1.5-fold precision over background.}
  \label{fig:sig5kbPH}
\end{figure}

\begin{figure}[h!]
  \centerline{\includegraphics[width=1.4\linewidth]{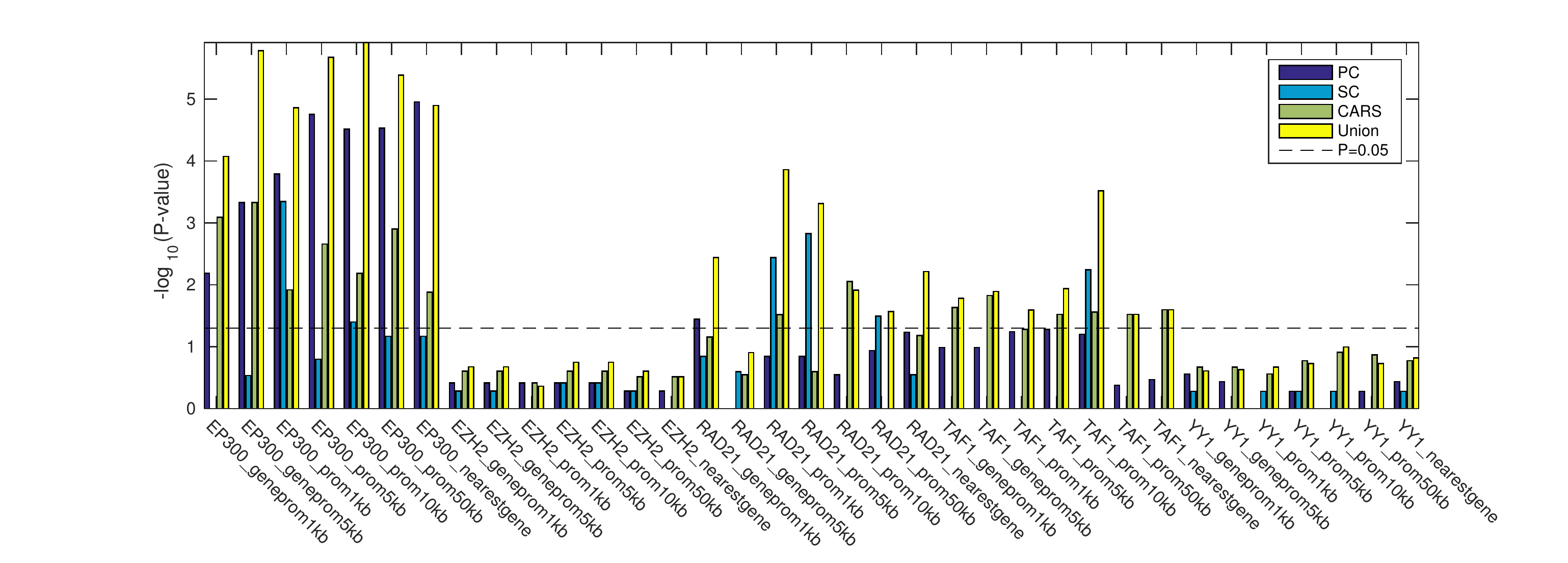}}
  \caption{Functional target set significance (hypergeometric P-value) using quantitative (sum of peak heights) ChIP-seq data predicted by each of the correlation methods for all peak-to-gene models at a predicted 1.5-fold precision over background.}
  \label{fig:sigAllPH}
\end{figure}

%% CpG partitioning:
\begin{figure}[h!]
  \centerline{\includegraphics[width=1.4\linewidth]{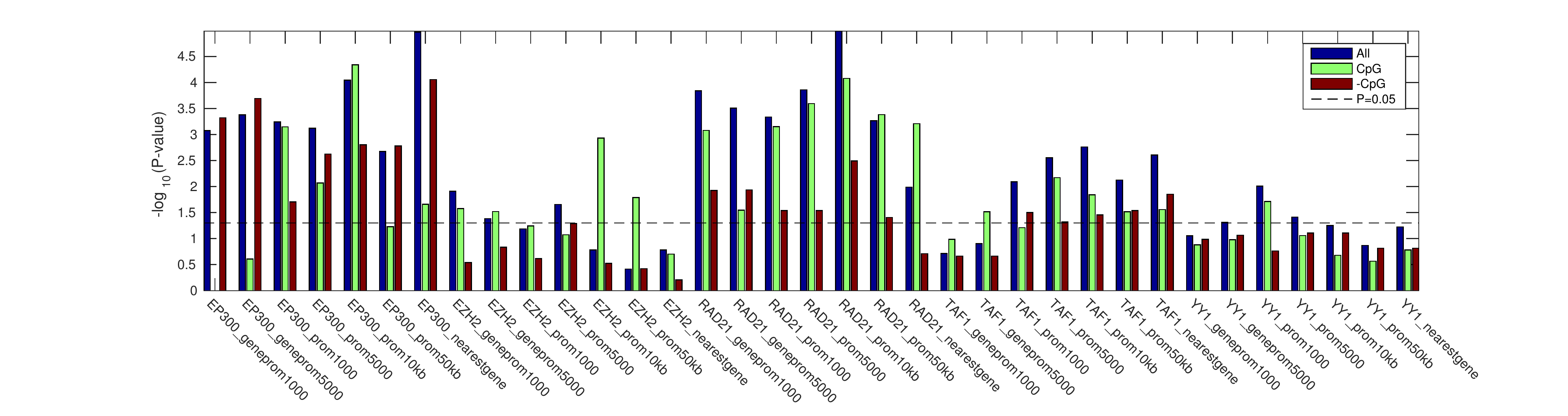}}
  \caption{Functional target set significance (hypergeometric P-value), using CpG partitioned datasets, predicted by each of the correlation methods for all peak-to-gene models at a predicted 1.5-fold precision over background. We show significance using all genes (blue), only CpG-rich promoters (green), and only CpG-depleted promoters (red).}
  \label{fig:CpGresults}
\end{figure}

\begin{figure}[h!]
  \centerline{\includegraphics[width=1.4\linewidth]{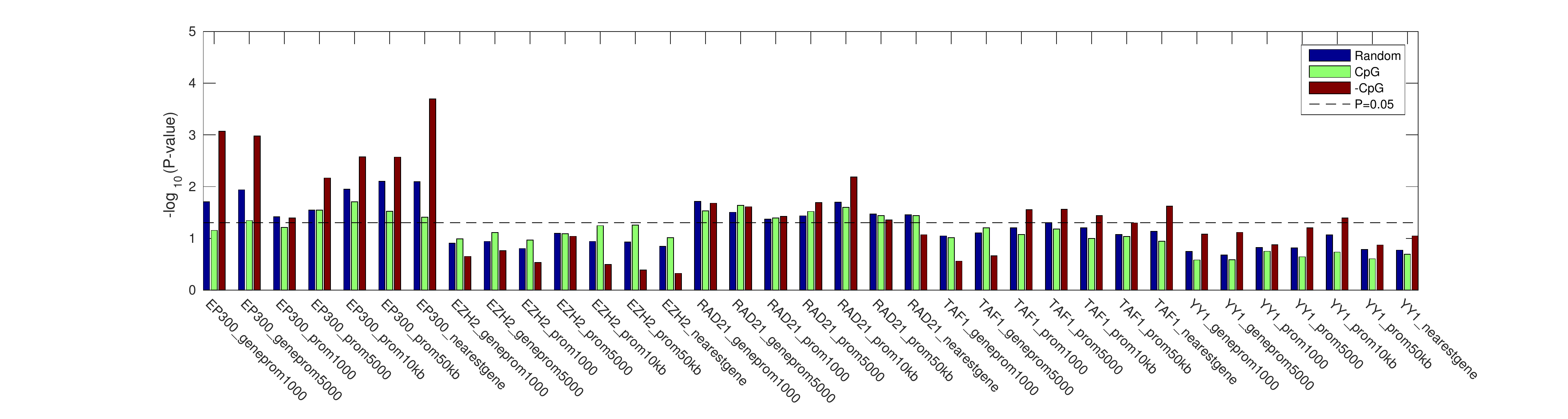}}
  \caption{Functional target set significance (hypergeometric P-value), using CpG partitioned datasets, each of 1000 randomly selected genes, predicted by each of the correlation methods for all peak-to-gene models at a predicted 1.5-fold precision over background. We show significance using all genes (blue), only CpG-rich promoters (green), and only CpG-depleted promoters (red). All results are the average of 100 random samples of 1000 genes per set.}
  \label{fig:CpGresults}
\end{figure}

\clearpage

\begin{table*}                             
  \centering
  \begin{tabular}{|ll|ccc|cc|}
    \hline
    & & \multicolumn{3}{c|}{Binding} &  \multicolumn{2}{c|}{MB} \\
    TF & Model & Background & Bound & p & n & p \\
    \hline                       
    EP300 & geneprom1000 & 0.385 & 0.424 & \textbf{0.000858} & 0 & 1 \\
    & geneprom5000 & 0.385 & 0.425 & \textbf{0.00048} & 0 & 1 \\
    & prom1000 & 0.385 & 0.394 & 0.347 & 0 & 1 \\
    & prom5000 & 0.385 & 0.413 & 0.0512 & 0 & 1 \\
    & prom10kb & 0.385 & 0.426 & \textbf{0.00679} & 23 & \textbf{0.0243} \\
    & prom50kb & 0.385 & 0.445 &\textbf{ 2.97e-06} & 86 & \textbf{0.000156} \\
    & nearestgene & 0.385 & 0.447 & \textbf{1.03e-07} & 29 & \textbf{0.0218} \\
    EZH2 & geneprom1000 & 0.216 & 0.186 & 0.923 & 0 & 1 \\
    & geneprom5000 & 0.216 & 0.197 & 0.827 & 0 & 1 \\
    & prom1000 & 0.216 & 0.178 & 0.877 & 0 & 1 \\
    & prom5000 & 0.216 & 0.216 & 0.529 & 0 & 1 \\
    & prom10kb & 0.216 & 0.21 & 0.607 & 0 & 1 \\
    & prom50kb & 0.216 & 0.212 & 0.59 & 0 & 1 \\
    & nearestgene & 0.216 & 0.205 & 0.72 & 0 & 1 \\
    RAD21 & geneprom1000 & 0.38 & 0.398 & \textbf{0.000132} & 18 & \textbf{0.0392} \\
    & geneprom5000 & 0.38 & 0.4 & \textbf{5.55e-06} & 19 & 0.0618 \\
    & prom1000 & 0.38 & 0.391 & 0.173 & 6 & 0.152 \\
    & prom5000 & 0.38 & 0.415 & \textbf{3.7e-06} & 12 & 0.125 \\
    & prom10kb & 0.38 & 0.423 & \textbf{6.05e-10} & 47 & \textbf{0.00509} \\
    & prom50kb & 0.38 & 0.405 & \textbf{3.08e-08} & 0 & 1 \\
    & nearestgene & 0.38 & 0.402 & \textbf{9.34e-08} & 1 & 0.38 \\
    TAF1 & geneprom1000 & 0.236 & 0.229 & 0.995 & 4 & 0.239 \\
    & geneprom5000 & 0.236 & 0.229 & 0.995 & 4 & 0.239 \\
    & prom1000 & 0.236 & 0.23 & 0.938 & 0 & 1 \\
    & prom5000 & 0.236 & 0.231 & 0.947 & 13 & 0.173 \\
    & prom10kb & 0.236 & 0.229 & 0.977 & 0 & 1 \\
    & prom50kb & 0.236 & 0.23 & 0.953 & 4 & 0.239 \\
    & nearestgene & 0.236 & 0.232 & 0.924 & 4 & 0.239 \\
    YY1 & geneprom1000 & 0.311 & 0.308 & 0.835 & 0 & 1 \\
    & geneprom5000 & 0.311 & 0.307 & 0.893 & 0 & 1 \\
    & prom1000 & 0.311 & 0.305 & 0.919 & 0 & 1 \\
    & prom5000 & 0.311 & 0.309 & 0.712 & 5 & \textbf{0.0349} \\
    & prom10kb & 0.311 & 0.31 & 0.57 & 14 & 0.109 \\
    & prom50kb & 0.311 & 0.308 & 0.756 & 3 & 0.229 \\
    & nearestgene & 0.311 & 0.308 & 0.791 & 4 & 0.367 \\
    \hline
  \end{tabular}
  \caption{Overlap between various predicted and known functional TF-target sets for ENCODE data. Binding: the ratio of differentially expressed genes among all 8,872 reference genes (Background) and among genes bound by the TF in the given peak-to-gene model (Bound), and the hypergeometric overlap \pval\ (p). Multiple Bind: gene sets predicted by a threshold on the number of peaks with a 1.5-fold increase in ratio of differentially expressed genes compared to the background, showing the number of genes (n) and hypergeometric overlap \pval\ (p). All subset sizes refer to the number of 8,872 reference genes exceeding the threshold. Significant \pval s ($<0.05$) are indicated in bold}
  \label{table:summary1}                         
\end{table*}

\begin{table*}                             
  \centering
  \begin{tabular}{|ll|cc|cc|cc|cc|}
    \hline
    & & \multicolumn{2}{c|}{PC} & \multicolumn{2}{c|}{SC} & \multicolumn{2}{c|}{CARS} & \multicolumn{2}{c|}{Union}\\
    TF & Model & n & p & n & p & n & p & n & p \\
    \hline    
    EP300 & geneprom1000 & 41 & \textbf{0.00709} & 0 & 1 & 29 & \textbf{0.0218} & 63 & \textbf{0.00084} \\
    & geneprom5000 & 43 & \textbf{0.00681} & 0 & 1 & 43 & \textbf{0.00681} & 78 & \textbf{0.000418} \\
    & prom1000 & 25 & \textbf{0.0236} & 17 & 0.0715 & 25 & \textbf{0.0236} & 58 & \textbf{0.000575} \\
    & prom5000 & 12 & 0.132 & 12 & 0.132 & 51 & \textbf{0.00246} & 69 & \textbf{0.000754} \\
    & prom10kb & 45 & \textbf{0.00653} & 25 & \textbf{0.0236} & 43 & \textbf{0.00681} & 93 & \textbf{9.07e-05} \\
    & prom50kb & 24 & \textbf{0.0381} & 8 & 0.151 & 31 & \textbf{0.0209} & 59 & \textbf{0.00211} \\
    & nearestgene & 119 & \textbf{1.07e-05} & 0 & 1 & 32 & \textbf{0.0131} & 138 & \textbf{1.08e-05} \\
    EZH2 & geneprom1000 & 6 & 0.383 & 3 & 0.518 & 92 & \textbf{0.00889} & 94 & \textbf{0.0123} \\
    & geneprom5000 & 6 & 0.383 & 6 & 0.383 & 46 & 0.0546 & 52 & \textbf{0.0415} \\
    & prom1000 & 3 & 0.518 & 0 & 1 & 46 & 0.0546 & 47 & 0.0648 \\
    & prom5000 & 9 & 0.304 & 9 & 0.304 & 61 & \textbf{0.0277} & 71 & \textbf{0.0221} \\
    & prom10kb & 3 & 0.518 & 3 & 0.518 & 9 & 0.304 & 14 & 0.166 \\
    & prom50kb & 0 & 1 & 6 & 0.383 & 0 & 1 & 6 & 0.383 \\
    & nearestgene & 3 & 0.518 & 5 & 0.295 & 9 & 0.304 & 14 & 0.166 \\
    RAD21 & geneprom1000 & 35 & \textbf{0.0162} & 3 & 0.323 & 61 & \textbf{0.00156} & 91 & \textbf{0.000145} \\
    & geneprom5000 & 0 & 1 & 6 & 0.152 & 70 & \textbf{0.000826} & 76 & \textbf{0.000311} \\
    & prom1000 & 26 & \textbf{0.0322} & 31 & \textbf{0.0181} & 47 & \textbf{0.00509} & 92 & \textbf{0.00046} \\
    & prom5000 & 21 & 0.058 & 56 & \textbf{0.00267} & 45 & \textbf{0.00534} & 108 & \textbf{0.000138} \\
    & prom10kb & 28 & \textbf{0.0304} & 34 & \textbf{0.0107} & 71 & \textbf{0.000532} & 116 & \textbf{1.03e-05} \\
    & prom50kb & 35 & \textbf{0.0162} & 54 & \textbf{0.00281} & 8 & 0.144 & 86 & \textbf{0.000544} \\
    & nearestgene & 0 & 1 & 22 & \textbf{0.0358} & 14 & 0.115 & 36 & \textbf{0.0103} \\
    TAF1 & geneprom1000 & 8 & 0.288 & 0 & 1 & 14 & 0.219 & 17 & 0.194 \\
    & geneprom5000 & 8 & 0.288 & 0 & 1 & 19 & 0.139 & 22 & 0.126 \\
    & prom1000 & 50 & \textbf{0.0328} & 18 & \textbf{0.0133} & 11 & 0.25 & 67 & \textbf{0.00809} \\
    & prom5000 & 47 & \textbf{0.0358} & 32 & 0.0552 & 70 & \textbf{0.0149} & 130 & \textbf{0.0028} \\
    & prom10kb & 0 & 1 & 36 & \textbf{0.0293} & 64 & \textbf{0.0176} & 100 & \textbf{0.00173} \\
    & prom50kb & 2 & 0.417 & 28 & 0.103 & 59 & \textbf{0.0253} & 87 & \textbf{0.00748} \\
    & nearestgene & 5 & 0.338 & 50 & \textbf{0.0328} & 62 & \textbf{0.0232} & 112 & \textbf{0.00247} \\
    YY1 & geneprom1000 & 10 & 0.169 & 0 & 1 & 6 & 0.275 & 16 & \textbf{0.0886} \\
    & geneprom5000 & 10 & 0.169 & 0 & 1 & 12 & 0.135 & 22 & \textbf{0.0489} \\
    & prom1000 & 47 & \textbf{0.0166} & 9 & 0.112 & 4 & 0.367 & 53 & \textbf{0.00983} \\
    & prom5000 & 8 & 0.213 & 17 & 0.123 & 8 & 0.213 & 29 & \textbf{0.0386} \\
    & prom10kb & 0 & 1 & 0 & 1 & 25 & 0.0563 & 25 & 0.0563 \\
    & prom50kb & 0 & 1 & 0 & 1 & 12 & 0.135 & 12 & 0.135 \\
    & nearestgene & 4 & 0.367 & 7 & 0.14 & 10 & 0.169 & 20 & 0.0594 \\
    \hline
  \end{tabular}
  \caption{Overlap between various predicted and known functional TF-target sets for ENCODE data. Pearson Correlation (PC), Spearman Correlation (SC), Combined Angle Ratio Statistic (CARS), and Union: gene sets predicted by a threshold on the correlation score, for each method respectively and for the union of those sets, with a 1.5-fold increase in ratio of differentially expressed genes compared to the background, showing the number of genes (n) and hypergeometric overlap \pval\ (p). All subset sizes refer to the number of 8,872 reference genes exceeding the threshold. Significant \pval s ($<0.05$) are indicated in bold}
  \label{table:summary2}                         
\end{table*}  

\begin{table*}                             
  \centering
  \begin{tabular}{|ll|ccc|cc|}
    \hline
    & & \multicolumn{3}{c|}{Binding} &  \multicolumn{2}{c|}{MB} \\
    TF & Model & Background & Bound & p & n & p \\
    \hline                       
    Per2 & geneprom1kb & 0.281 & 0.281 & 0.486 & 1 & 0.281 \\
    & geneprom5kb & 0.281 & 0.282 & 0.266 & 11 & 0.17 \\
    & prom1kb & 0.268 & 0.269 & 0.397 & 0 & 1 \\
    & prom5kb & 0.272 & 0.277 & \textbf{0.0414} & 0 & 1 \\
    & prom10kb & 0.28 & 0.284 & 0.0535 & 30 & \textbf{0.0217} \\
    & prom50kb & 0.279 & 0.281 & 0.141 & 21 & 0.102 \\
    & nearestgene & 0.278 & 0.28 & 0.101 & 2 & 0.478 \\
    Cry1 & geneprom1kb & 0.0126 & 0.0126 & 0.734 & 939 & \textbf{0.0196} \\
    & geneprom5kb & 0.0129 & 0.0129 & 0.726 & 580 & 0.0652 \\
    & prom1kb & 0.0109 & 0.0108 & 0.796 & 70 & \textbf{0.0393} \\
    & prom5kb & 0.0123 & 0.0124 & 0.644 & 365 & 0.0752 \\
    & prom10kb & 0.0124 & 0.0125 & 0.562 & 234 & 0.068 \\
    & prom50kb & 0.0127 & 0.0128 & 0.439 & 213 & 0.132 \\
    & nearestgene & 0.0127 & 0.0127 & 0.7 & 916 & \textbf{0.035} \\
    Cry2 & geneprom1kb & 0.014 & 0.0136 & 0.897 & 1040 & \textbf{0.0177} \\
    & geneprom5kb & 0.014 & 0.0136 & 0.896 & 185 & 0.257 \\
    & prom1kb & 0.0128 & 0.0121 & 0.903 & 171 & 0.171 \\
    & prom5kb & 0.013 & 0.0125 & 0.89 & 106 & 0.157 \\
    & prom10kb & 0.014 & 0.0134 & 0.932 & 24 & 0.287 \\
    & prom50kb & 0.0149 & 0.0146 & 0.858 & 103 & 0.198 \\
    & nearestgene & 0.0148 & 0.0147 & 0.701 & 378 & 0.103 \\
    \hline
  \end{tabular}
  \caption{Overlap between various predicted and known functional TF-target sets for mouse circadian data. Binding: the ratio of differentially expressed genes among all 8,872 reference genes (Background) and among genes bound by the TF in the given peak-to-gene model (Bound), and the hypergeometric overlap \pval\ (p). Multiple Bind: gene sets predicted by a threshold on the number of peaks with a 1.5-fold increase in ratio of differentially expressed genes compared to the background, showing the number of genes (n) and hypergeometric overlap \pval\ (p). Significant \pval s ($<0.05$) are indicated in bold}
  \label{table:summaryC1}                         
\end{table*}

\begin{table*}                             
  \centering
  \begin{tabular}{|ll|cc|cc|cc|cc|}
    \hline
    & & \multicolumn{2}{c|}{PC} & \multicolumn{2}{c|}{SC} & \multicolumn{2}{c|}{CARS} & \multicolumn{2}{c|}{Union}\\
    TF & Model & n & p & n & p & n & p & n & p \\
    \hline    
    Per2 & geneprom1kb & 2 & 0.483 & 10 & 0.119 & 0 & 1 & 10 & 0.119 \\
    & geneprom5kb & 0 & 1 & 10 & 0.119 & 0 & 1 & 10 & 0.119 \\
    & prom1kb & 4 & 0.292 & 2 & 0.464 & 0 & 1 & 6 & 0.198 \\
    & prom5kb & 17 & 0.153 & 3 & 0.182 & 0 & 1 & 19 & 0.117 \\
    & prom10kb & 0 & 1 & 9 & 0.225 & 0 & 1 & 9 & 0.225 \\
    & prom50kb & 0 & 1 & 0 & 1 & 0 & 1 & 0 & 1 \\
    & nearestgene & 0 & 1 & 0 & 1 & 0 & 1 & 0 & 1 \\
    Cry1 & geneprom1kb & 2164 & \textbf{0.00039} & 1728 & \textbf{0.000496} & 528 & 0.124 & 2432 & \textbf{0.000524} \\
    & geneprom5kb & 1807 & \textbf{0.00239} & 1807 & \textbf{0.00239} & 619 & 0.0959 & 2288 & \textbf{0.000677} \\
    & prom1kb & 856 & 0.0549 & 61 & 0.491 & 367 & 0.203 & 1015 & 0.0509 \\
    & prom5kb & 1026 & \textbf{0.0301} & 702 & 0.0778 & 486 & 0.137 & 1298 & \textbf{0.0117} \\
    & prom10kb & 268 & 0.238 & 45 & 0.107 & 536 & 0.121 & 691 & 0.138 \\
    & prom50kb & 420 & 0.16 & 1204 & \textbf{0.0207} & 473 & 0.142 & 1493 & \textbf{0.012} \\
    & nearestgene & 787 & \textbf{0.0671} & 1393 & \textbf{0.0101} & 577 & 0.111 & 1760 & \textbf{0.00607} \\
    Cry2 & geneprom1kb & 1570 & \textbf{0.00144} & 566 & 0.087 & 54 & 0.536 & 1609 & \textbf{0.00238} \\
    & geneprom5kb & 1670 & \textbf{0.00103} & 608 & \textbf{0.0356} & 95 & 0.385 & 1727 & \textbf{0.000851} \\
    & prom1kb & 622 & 0.0659 & 229 & 0.162 & 0 & 1 & 622 & 0.0659 \\
    & prom5kb & 974 & \textbf{0.0201} & 422 & \textbf{0.0368} & 512 & 0.111 & 1144 & \textbf{0.00134} \\
    & prom10kb & 1145 & \textbf{0.0095} & 427 & \textbf{0.0287} & 572 & 0.0873 & 1339 & \textbf{0.00123} \\
    & prom50kb & 44 & 0.485 & 0 & 1 & 760 & \textbf{0.0484} & 773 & 0.0555 \\
    & nearestgene & 1352 & \textbf{0.00667} & 774 & \textbf{0.0289} & 135 & 0.322 & 1468 & \textbf{0.00298} \\
    \hline
  \end{tabular}
  \caption{Overlap between various predicted and known functional TF-target sets for mouse circadian data. Pearson Correlation (PC), Spearman Correlation (SC), Combined Angle Ratio Statistic (CARS), and Union: gene sets predicted by a threshold on the correlation score, for each method respectively and for the union of those sets, with a 1.5-fold increase in ratio of differentially expressed genes compared to the background, showing the number of genes (n) and hypergeometric overlap \pval\ (p). Significant \pval s ($<0.05$) are indicated in bold}
  \label{table:summaryC2}   
\end{table*}

%% Data sources
\begin{table}
  \centering
  \begin{tabular}{|l|l|l|}
    \hline
    Data & TFs & Source \\
    \hline\hline
    \multicolumn{3}{|l|}{Human} \\
    \hline\hline
    ChIP-seq & EP300 &  \\
     & EZH2 &  \\
     & RAD21 &  \\
     & TAF1 & https://genome.ucsc.edu/ENCODE/dataMatrix/encodeChipMatrixHuman.html \\
     & YY1 &  \\
     & CEBPB &  \\
     & MYC &  \\
     & REST &  \\
    \hline
    RNA-seq & EP300 &  \\
     & EZH2 &  \\
     & RAD21 &  \\
     & TAF1 & https://genome.ucsc.edu/ENCODE/dataMatrix/encodeDataMatrixHuman.html \\
     & YY1 &  \\
     & CEBPB &  \\
     & MYC &  \\
     & REST &  \\
    \hline
    Knockout & EP300 &  \\
     & EZH2 &  \\
     & RAD21 &  Cusanovich et al.\ (2014)\\
     & TAF1 &  \\
     & YY1 &  \\
    \hline\hline
    \multicolumn{3}{|l|}{Mouse} \\
    \hline\hline
    ChIP-seq & Bmal1 &  \\
    \& & Clock &  \\
    RNA-seq & Cry1 & Koike et al.\ (2012) \\
     & Cry2 &  \\
     & Per1 &  \\
     & Per2 &  \\
    \hline
    Knockout & Per2 & http://www.ncbi.nlm.nih.gov/geo/query/acc.cgi?acc=GSE30139 \\
    \hline
  \end{tabular}
  \caption{Data sources.}
  \label{tab:data-sources}
\end{table}

\end{document}